\documentclass[a4paper, 11pt]{article}
\usepackage{amsfonts, amssymb, amsmath, amsthm, graphicx}
\newtheorem{thm}{Theorem}

\newtheorem{defn}[thm]{Definition}
\newtheorem{coly}[thm]{Corollary}
\newtheorem{lemma}[thm]{Lemma}

\newcommand{\ket}[1]{{|#1\rangle}}

\newcommand{\B}{\{0,1\}}

\newcommand{\pfstart}{\begin{proof}} 
\newcommand{\pfend}{\end{proof}} 
\newcommand{\s}[1]{\left(#1\right)}

\begin{document}
\title{Quantum query complexity of subgraph containment with constant-sized certificates}
\author{Yechao Zhu\thanks{Queens' College, Cambridge, CB3 9ET, UK; work done while the author was a summer student at Centre for Quantum Information and Foundations, Department of Applied Mathematics and Theoretical Physics, University of Cambridge; yz342@cam.ac.uk}}
\date{\vspace*{-.7cm}}

\maketitle

\begin{abstract}
We study the quantum query complexity of constant-sized subgraph containment. Such problems include determining whether a $ n $-vertex graph contains a triangle, clique or star of some size. For a general subgraph $ H $ with $ k $ vertices, we show that $ H $ containment can be solved with quantum query complexity $ O\left(n^{2-\frac{2}{k}-g(H)}\right) $, with $ g(H) $ a strictly positive function of $ H $. This is better than $ \tilde{O}\s{n^{2-2/k}} $ by Magniez \textit{et al}. This result is obtained in the learning graph model of Belovs.
\end{abstract}

\section{Introduction}
The oracle model of quantum computing and its associated notion of quantum query complexity imply that quantum computers may be much more powerful than classical computers and have provided lots of incentive for people to study this area.

The oracle model is one in which the input is given as an oracle and the only means to gain knowledge about the input is by asking queries to the oracle. The query complexity of an algorithm is the number of queries that it makes. Many famous quantum algorithms are based on this model, such as Deutsch-Jozsa algorithm \cite{Deutsch92}, Grover's search algorithm \cite{Grover96}, and the period-finding part of Shor's factoring algorithm \cite{Shor97}.

One particular area that people focus on is lower bounds of quantum query complexities. Beals \textit{et al} used the polynomial method and showed that the exponential quantum speed-up obtained for partial functions cannot be obtained for any total function and showed such query complexities are polynomially related \cite{Beals01}. Ambainis proposed the quantum adversary method \cite{Ambainis02} and argued that if the function runs on a superposition of inputs, the algorithm part and oracle part will achieve some entanglement and the number of queries that are needed for this entanglement implies a lower bound on query complexity.

Another area is to develop new tools for the construction of quantum query algorithms. Discrete time quantum walk was introduced by Meyer and Watrous\cite{Meyer96, Meyer96a, Watrous01}. Since then, quantum walks have been applied to Grover search \cite{Shenvi03}, element distinctness \cite{Ambainis07} and matrix product verification \cite{Buhrman06}.

Many such quantum algorithms have been shown to be optimal, i.e. the query complexity of the algorithm matches the lower bound of the problem. For example, in the element distinctness problem, one is given input elements $ x_1, \cdots, x_n \in \{0, 1, \cdots, m-1\} $ and is asked if there exist $ i,j $ such that $ x_i=x_j $. Aaronson and Shi gave the lower bound $ \Omega\s{n^{2/3}} $ \cite{Aaronson04}. Ambainis gave a quantum algorithm using a quantum walk on the Johnson graph, with query complexity $ O\s{n^{2/3}} $, matching the lower bound \cite{Ambainis07}.

Despite all these work, there are still many problems in which an optimal algorithm has not been obtained. One such problem is the triangle problem, which is to determine if a graph contains a triangle. Buhrman \textit{et al}. first studied the problem and gave an algorithm with quantum query complexity $ O\s{n+\sqrt{nm}} $, for graphs with $ n $ vertices and $ m $ edges \cite{Buhrman05}.  This result uses $ O\s{n^{3/2}} $ queries when $ m=\Theta(n^2) $. This is then improved by Magniez, Santha, and Szegedy to $ \tilde{O}\s{n^{1.3}} $ \cite{Magniez07}, and then improved by Magniez, Nayak, Roland and Santha to $ O\s{n^{1.3}} $ \cite{Magniez07a}. However, the best known lower bound remains $ \Omega(n) $ since it cannot be improved using the quantum adversary method due to a limitation on 1-certificate complexity \cite{Zhang05}.

Recently, a new computational model called span program was introduced to analyse quantum query complexities. It was shown to be equivalent to quantum query algorithms, up to a constant factor, by Reichardt\cite{Reichardt10a, Reichardt11}. Although this is a very important result, even for simple problems, it is difficult to come up with explicit span programs. The learning graph model introduced by Belovs provides a way of generating span programs \cite{Belovs11}. It can be used to analyse the query complexities of various search problems, which were previously solved by quantum walks. In particular, this model does not involve quantum walks and spectral analysis, but is yet very powerful. Belovs recovered the previous results on element distinctness and triangle problem and improved the quantum query complexity for the triangle problem to $ O\s{n^{35/27}} $.

In this paper, we explore the power of the learning graph further and study the quantum query complexity of subgraph containment, i.e. to determine if a graph contains a particular subgraph $ H $. This includes a large class of graph properties, such as determining if a graph contains a path or cycle of given length, a clique (complete subgraph) or a star of given size. It was solved by Magniez \textit{et al}. in $ \tilde{O}\s{n^{2-2/k}} $ queries, with $ k $ the number of vertices in $ H $ \cite{Magniez07}. However, similar to the triangle problem, the best known lower bound is again $ \Omega\s{n} $.

We firstly extend the results of the distinctness problem to $ k $-distinctness problem. We give a learning graph with query complexity $ O\left(n^{k/(k+1)}\right) $, matching a previous result of Ambainis \cite{Ambainis07}. This result was obtained independently by Belovs and Lee \cite{Belovs11a}. We then use it as a subroutine in constructing learning graphs for various graph containment problems. We first improve the query complexity of $ H $-subgraph (of $ k $ vertices) containment from  $ \tilde{O}\s{n^{2-2/k}} $ to $ O\s{n^{2-2/k}} $ (Theorem \ref{thm:containmentold}). Then we use another type of learning graph in \cite{Belovs11} and derive our main theorem. This reproduces Belovs' result in the case $ H $ is a triangle.\\
\textit{If the subgraph $ H $ has $ k \ge 3$ vertices, $ m+l $ edges, with $ l $ being the minimum degree of a vertex in $ H $, then $ H $-containment has quantum query complexity $ O\s{n^{2-2/k-g(H)}} $, with $ g(H)=\frac{2k-l-3}{k(l+1)(m+2)} $ is strictly positive (Theorem \ref{thm:containment}).}

\section{Preliminaries}
\subsection{Certificate complexity}
We work with functions having constant 1-certificate complexity in this paper. Consider a multivariable function $ f:[m]^n \to \B $. By $ [m] $, we denote the set $ \{0,1,\cdots,m-1\} $. For $ x\in f^{-1}\s{b} $, a $ b $-certificate is a sequence of variables in $ x $ that proves $ f\s{x}=b $. The $ b $-certificate complexity of $ f $, denoted by $ C^{(b)}(f) $, is $ \max_{x\in f^{-1}\s{b}} \{ $number of variables in the smallest $ b $-certificate of $ x\} $.

For $ x\in f^{-1}\s{1} $, $ x $ may have a few 1-certificates. For the problems that we are interested in, we will just choose one 1-certificate for each $ x\in f^{-1}\s{1} $. Suppose, some $ x\in f^{-1}\s{1} $ has a 1-certificate $ \s{x_{i_1}, x_{i_2}, \cdots, x_{i_k}} $, then we call $ \s{i_1, \cdots, i_k} $ \textit{marked elements}.

\subsection{Query Model}
We are interested in the query complexity of $ f $ in the standard query complexity model \cite{Buhrman02}. The query complexity of an algorithm is the number of queries needed for the algorithm to compute $ f $. The query complexity of $ f $ is the minimum query complexity over all algorithms computing $ f $. This is different from time complexity, which is the number of basic operations by an algorithm. Naturally, the time complexity of $ f $ is at least as large as its query complexity.

The input state is $ \ket{i,y,z} $, where $ i $  is the index, $ y\in [m] $, and $ \ket{z} $ is some ancilla state. A query $ O $ maps $ \ket{i,y,z} $ to $ \ket{i,y\oplus x_i,z} $ ($ \oplus $ denote addition modulo $ m $). A quantum computation with $ k $ queries would be a sequence of operations $ U_0, O, U_1, O, \cdots, O, U_k $, where $ U_i $'s are unitary transformations that do not depend on the input.

\section{Learning Graph Model}
\subsection{Definitions}
We define the learning graph following, mostly, \cite{Belovs11} (defined in \cite{Belovs11} as reduced learning graph). The learning graph model of computation is introduced for a function $ f: [m]^n \to \B$ with Boolean output.
\begin{defn}
A learning graph is a directed acyclic connected graph with vertices (sometimes called L-vertices) labelled by subsets of $ [n] $, the input indices. It starts from $ \emptyset $ and has arcs connecting $ S $ and $ S' $, where $ S\subseteq[n] $ and $ S\subset S'\subseteq [n] $. We call these arcs \textit{transitions}. The \textit{length} $ \ell(e) $ of the transition $ e $ is defined as $ |S' \setminus S| $.
\end{defn}

For each $ x\in f^{-1}\s{1} $, we can define a flow $ p_e\s{x} $ on transition $ e $. We can think of flows as probabilities and the process as a random walk on the learning graph with some probability of using each transition, defined by its flow. Hence, the flows originate from $ \emptyset $ (the source). The sum of flows over all transitions leaving $ \emptyset $ is 1. If the learning graph does not terminate at a vertex, then the flow through that vertex is preserved. For different input, the probability of using a transition is definitely different. Hence, the flows depend on the input and the marked elements.

Define the distance $ d\s{v} $ of a vertex as the number of transitions required to connect it to $ \emptyset $. A layer $ V_k $ is a collection of vertices which have the same distance $ k $, which we can define as the distance of the layer. A stage is the set of all transitions from $ V_k $ to $ V_{k+1} $. Naturally, the flows through all transitions in a stage have sum 1.

We will just describe the learning graph by the stages. Each stage can be thought of as accessing a superposition of all elements in $ S'\setminus S $, for all $ S\in V_k $, $ S'\in V_{k+1} $. The learning graph finishes once we have \textit{found} the marked elements according to the description. A transition is $ valid $ if it is used in the flow. The flows through each transition will be obvious from the description. Of course one may enumerate each valid transition and assign flows manually.

Stages represent our progress towards finding the marked elements. So, if we are on the right track according to the description, we should be using a valid transition at any stage. The probability that we are using a particular valid transition is its flow. We progress by making queries into the input (like in a quantum walk). Hence, we describe a stage as querying some elements. 

We illustrate this with a learning graph for the distinctness problem. If the function to be evaluated is $ f: [m]^5\to \B $ and the input variable $ x\in f^{-1}(1) $ has marked elements $ a $ and $ b $, we could have three stages.

\begin{table}[h]
\begin{tabular}{rp{13cm}}
\hline
1.& Query $ 2 $ elements (uniformly at random) different from $ a $ and $ b $.\\
2. & Query $ a $.\\
3. & Query $ b $.\\
\hline
\end{tabular}
\caption{Stages for the distinctness problem.}
\label{tbl:distinctness}
\end{table}

\begin{figure}[tbh]
\centering
\includegraphics[width=11cm]{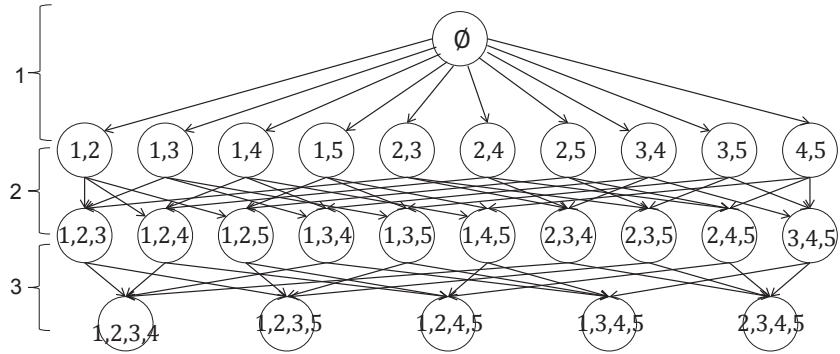}
\caption{The learning graph for the distinctness problem. Stages 1, 2 and 3 shown}
\label{fig:distinct0}
\end{figure}

Stage 1 includes all transitions starting from $ \emptyset $ and which queries two elements. A transition in Stage 1 is valid if it queries neither $ a $ nor $ b $. Out of the 10 transitions in Stage 1, 3 are valid. Hence, the flow through a valid transition in Stage 1 is $ \frac{1}{3} $. Stage 2 includes all transitions starting with a 2-element vertex and which queries one element. A transition in Stage 2 is valid if it queries $ a $ and the starting vertex does not include element $ b $.

\subsection{Using symmetry}
Denote the group that leaves the problem invariant by $ \mathbb{S} $. $ \mathbb{S} $ can be the full symmetry group on the input indices, in the case of the search problem and the distinctness problem. For graphs, $ \mathbb{S} $ is the symmetry group on the vertices.

$ \mathbb{S} $ acts on the L-vertices of the learning graph by permuting the input indices. Assume the learning graph stays invariant under the action of $ \mathbb{S} $. For each transition $ e $ from $ S $ to $ S' $, we can define a natural group action by $ \mathbb{S} $ on $ e $, in the sense that $ g\s{e} $ is the transition from $ g\s{S} $ to $ g\s{S'} $, where $ g\in \mathbb{S} $. This defines an equivalence relationship, as two transitions are equivalent if one can be obtained from another by an element of $ \mathbb{S} $. $ \mathbb{S} $ partitions the transitions in a stage into equivalence classes.

We define \textit{speciality} $ \tau(e) $ of a transition $ e $ as the ratio of the size of the equivalence class containing $ e $ to the number of valid transitions in it. It equals the inverse of the probability of obtaining a valid transition when a random element from $ \mathbb{S} $ is applied to $ e $.

Assume a set of flows $ \{p_e(x): e$ is a transition, $ x \in f^{-1}(1) \} $ have been fixed. If the flows through all valid and equivalent transitions on stage $ i $ are equal, and the expression of $ p_e(x) $ and $ \tau(e) $ do not depend on the input, then we say the learning graph is \textit{symmetric} on stage $ i $. Similarly, one can define equivalence and symmetry of vertices in a layer.

Define the average length $ L_i $ of stage $ i $ by $ \sum\limits_{e\in E_i}p_e(x)\ell(e) $. It will not depend on the input if the flow is symmetric. Let $ T_i $ denote the maximal speciality of a transition in stage $ i $.

The definition of complexity for a stage and a learning graph in \cite{Belovs11} is rather cumbersome and does not provide much insight. Hence, we shall just give the following theorems which are sufficient for most applications.

\begin{thm}[{\cite{Belovs11}}]
Assume the flow is symmetric on stage $ i $. Then, the complexity of stage $ i $ is at most $ L_i\sqrt{T_i} $.
\label{thm:symmetry}
\end{thm}

Intuitively, if we think in terms of quantum walks, $ L_i $ can be interpreted as the average number of queries needed for a transition and $ \sqrt{T_i} $ as the number of steps repeated to amplify the amplitude of the valid transitions, i.e. to boost the success probability. Hence, the complexity of the entire stage is at most $ L_i\sqrt{T_i} $.

\begin{thm}[{\cite{Belovs11}}]
\label{thm:summa}
For a learning graph with $ k $ stages and complexity $ C_i $ for each stage, one can build a learning graph of complexity $ O \s{\sum\limits_{i=1}^{k}C_i} $.
\end{thm}

The following theorem relates the complexity of the learning graph to the query complexity of the function.

\begin{thm}[{\cite{Belovs11}}]
For any learning graph of a function $ f: [m]^n \to \B $ with complexity $ C $, there exists a bounded error quantum query algorithm for the same function with complexity $ O(C\log m) $.
\end{thm}

This result is proved in \cite{Belovs11} via span programs \cite{Reichardt10}. For $ m $ fixed, the complexity reduces to $ O(C) $. For the subgraph containment problem that we are interested in, the input is the adjacency matrix representing the input graph and hence $ m=2 $. In \cite{Belovs11a}, a slightly different learning graph model is proposed and the result is improved to $ O\s{C} $ via the general adversary bound.

\subsection{Subroutines}
One way to simplify a learning graph is to introduce the concept of subroutines. It is an existing learning graph that can be appended to the vertex of another learning graph. Suppose we have a vertex $ S \subseteq [n] $ in a learning graph $ G $. One can treat $ S $ as the initial vertex of another problem with $ [n]\setminus S $ as the set of input indices. Let $ G' $ be a learning graph for this new problem. One can append $ G' $ after $ S $ to $ G $. Of course, to preserve the flow through $ S $, we have to multiply the flows in $ G' $ by $ p_S $, the in-flow at $ S $. We allow a \textit{subroutine stage} as the last stage in a reduced learning graph.

One can apply symmetry to the subroutine stage, as it is done for transitions. Suppose the subroutine stage starts from $ V_k $. Let $ \ell(v) $ be the complexity of the subroutine appended to vertex $ v\in V_k $. Define the \textit{average complexity} $ L $ of the subroutine stage as $ \sum_{v\in V_k}p_v\ell(v) $. Let $ T $ denote the maximal speciality of a vertex in $ V_k $.

\begin{thm}[{\cite{Belovs11}}]
Suppose the flow is symmetric for vertex set $ V_k $. Then, the complexity of the subroutine stage is $ L\sqrt{T} $.
\label{thm:subroutine}
\end{thm}

\section{Generalisation of the Distinctness Problem}
Consider the element $ k $-distinctness problem. The function outputs 1 iff there is a set of $ k $ identical elements among $ n $ input elements. This is a natural generalisation of the distinctness problem. For an input $ x \in f^{-1}(1) $, let $ a_1, \cdots, a_k $ be the set of marked elements. Here we give a learning graph with complexity $ O\left(n^{k/(k+1)}\right) $. This matches the result in \cite{Ambainis07}.

\begin{table}[htb]
\begin{tabular}{rp{13cm}}
\hline
1.& Query $ r-k $ elements (uniformly randomly) different from $ a_1,\cdots a_k $.\\
2. & Query $ a_1 $.\\
3. & Query $ a_2 $.\\
$ \vdots $\\
$ k $. & Query $ a_{k-1} $.\\
$ k+1 $. & Query $ a_k $.\\
\hline
\end{tabular}
\caption{Stages for the $ k $-distinctness problem.}
\label{tbl:kdistinctness}
\end{table}

\begin{table}[htb]
\centering 
\begin{tabular}{|r|cccccc|}
\hline
Stage & 1 & 2 & 3 & $ \cdots $ & $ k $ & $ k+1 $ \\
\hline
Speciality & 1 & $n$ & $n^2/r$ & $ \cdots $ &$ n^{k-1}/r^{k-2} $ & $ n^{k}/r^{k-1} $ \\
Length & $r$ & 1 & 1 & $ \cdots $ & 1 & 1 \\
\hline
\end{tabular}\caption{Parameters (up to a constant factor) of the stages in the learning graph described by Table \ref{tbl:kdistinctness}}
\label{tbl:kdistinctnessParam}
\end{table}

Let's look at Stage $ j $. If a transition queries some element $ b $, by applying a random permutation, the probability that $ b $ is mapped to $ a_{j-1} $ is $ 1/n $. For $ 1\le m\le j-2 $, the probability that $ a_m $ is in the L-vertex from which the transition originates is $ O\s{r/n} $. Moreover, all transitions in this stage are obviously equivalent. Hence, the speciality of this stage is $ O\s{n^{j-1}/r^{j-2}} $. Its length is obviously 1 since every transition queries 1 element.

There are $ \binom{n-k}{r-k} $ valid transitions in Stage 1. Hence, the flow in each valid transition is $ {\binom{n-k}{r-k}}^{-1} $. Each valid transition in Stage 1 is followed by only one valid transition in Stage 2 and so forth, so the flow in each valid transition in any stage is $ {\binom{n-k}{r-k}}^{-1} $. Obviously the expression of the flow does not depend on the input. Therefore all stages are symmetric.

By Theorem \ref{thm:symmetry} and \ref{thm:summa}, the total complexity is
\begin{center}
$ O(r+\sum\limits_{j=1}^{k}\sqrt{n^j/r^{j-1}}) $.
\end{center}

The optimal value is $ O\left(n^{k/(k+1)}\right) $, attained when $ r=n^{k/(k+1)} $.
Note that the complexities of Stages 2 to $ k $ are dominated by the complexity of Stage $ k+1 $. We will encounter this issue again in other problems.

\section{Subgraph containment problem}
Here we give a learning graph for $ k $-clique containment. A $ k $-clique is a complete graph on $ k $ vertices. This is a natural extension of the first learning graph for the triangle problem in \cite{Belovs11}. For an input graph in $ f^{-1}(1) $, there exists a $ k $-clique in the graph. Denote the vertices of this $ k $-clique by $ a_1, \cdots, a_k $.
\begin{thm}
$ k $-clique containment has quantum query complexity $ O \left(n^{2-2/k}\right) $.
\label{thm:kclique}
\end{thm}

\pfstart
\begin{table}[htb]
\centering
\begin{tabular}{rp{13cm}}
\hline
1. & Query a complete subgraph on $ r-k+1 $ vertices that does not contain vertices $ a_1,\cdots a_k $.\\
2. & Query all edges connecting $ a_1 $ to the subgraph.\\
3. & Query all edges connecting $ a_2 $ to the subgraph (including $ a_1 $).\\
$ \vdots $\\
$ k $. & Query all edges connecting $ a_{k-1} $ to the subgraph (including $ a_1, \cdots a_{k-2} $).\\
$ k+1 $. & Use a subroutine from Table \ref{tbl:kdistinctness} to query edges $ a_1a_k, \cdots a_{k-1}a_k $ out of all edges connecting $ a_k $ to the subgraph\\
\hline
\end{tabular}
\caption{Stages for the $ k $-clique containment}
\label{tbl:kclique}
\end{table}
\begin{table}[htb]
\centering 
\begin{tabular}{|r|cccccc|}
\hline
Stage & 1 & 2 & 3 & $ \cdots $ & $ k $ & $ k+1 $  \\
\hline
Speciality & 1 & $n$ & $n^2/r$ & $ \cdots $ &$ n^{k-1}/r^{k-2} $ & $ n^{k}/r^{k-1} $\\
Length/Complexity & $r^2$ & $ r $ & $ r $ & $ \cdots $ & $ r $ & $ r^{(k-1)/k} $ \\
\hline
\end{tabular}\caption{Parameters (up to a constant factor) of the stages in the learning graph described by Table \ref{tbl:kclique}}
\label{tbl:kcliqueParam}
\end{table}

After Stage $ k $, our task becomes to query edges $ a_1a_k, \cdots, a_{k-1}a_k $ out of the $ r $ edges connecting $ a_k $ to the complete subgraph of $ r $ vertices. If we think of the $ r $ edges as input variables, and $ a_1a_k, \cdots, a_{k-1}a_k $ as the marked elements, then this is essentially a $ k $-distinctness type problem. Therefore, we use a subroutine from the $ k $-distinctness problem to complete the last stage.

It can be checked that the obvious uniform flow satisfies the symmetry requirement. Hence, by Theorem \ref{thm:symmetry}, \ref{thm:summa} and \ref{thm:subroutine}, the total complexity is \begin{center}
$ O \left(r^2+r\sum\limits_{i=1}^{k-1}\sqrt{\frac{n^i}{r^{i-1}}}+r^{(k-1)/k}\sqrt{\frac{n^k}{r^{k-1}}} \right)$.
\end{center}

Notice that Stage $ k $ dominates all other stages except stages 1 and $ k+1 $. So the complexity is simplified to $ O \left(r^2+r\sqrt{\frac{n^{k-1}}{r^{k-2}}}+r^{(k-1)/k}\sqrt{\frac{n^k}{r^{k-1}}}\right) $, which is optimised to $ O \left(n^{2-2/k}\right) $ when $ r=n^{1-1/k} $.
\pfend

Since the last stage is a subroutine stage and its speciality is that of the vertex instead of a transition, we can relax this stage not to query all of  $ \s{a_1a_k, \cdots a_{k-1}a_k} $. Also, the previous $ k $ stages do not depend on the property of cliques. Hence, this learning graph can be generalised for any subgraph $ H $ with $ k $ vertices. Suppose $ a_k $ is a vertex in $ H $ with degree $ m $. Then we can use a subroutine to query $ m $ edges.

The complexity of the subroutine is $O \left( r^{1-1/m}\right) $, which is smaller than the corresponding one in $ k $-clique containment. Hence, the total complexity remains the same.

\begin{thm}
If $ H $ is a subgraph with $ k\geq 3 $ vertices, then $ H $ containment has quantum query complexity $ O \left(n^{2-2/k}\right) $.
\label{thm:containmentold}
\end{thm}

The second learning graph for the triangle problem in \cite{Belovs11} offers more freedom, since it queries random subgraphs rather than complete subgraphs, and queries the edges in the 1-certificate in a separate stage. Together with the technique we developed above, we constructed another learning graph for the general subgraph containment with small 1-certificates.

Before that, we give an important result which is implicitly used in \cite{Belovs11}. For a stage of complexity $ C $, with flows $ p_e\s{x} $ defined for each transition $ e $, for each input $ x $. Let $ E $ be the set of all transitions in this stage. $ E'\s{x}\subset E $. $ \sum\limits_{e\in E'\s{x}}p_e\s{x}=p\s{x} $. We can modify the flow this way:
\begin{itemize}
\item For $ e\in E'\s{x} $, $ p_e'\s{x}=\frac{p_e\s{x}}{p\s{x}} $.
\item For $ e\notin E'\s{x} $, $ p_e'\s{x}=0 $.
\end{itemize}
In this case, the sum of flow in this stage is still 1.
\begin{lemma}
\label{thm:modifyflow}
Suppose $ \max_{x\in f^{-1}(1)}\frac{1}{p\s{x}}<K $ for some constant $ K $, then the new complexity of the stage is $ O\s{C} $.
\end{lemma}

For a graph H of $ k \ge 3$ vertices, $ m+l $ edges, with $ l $ being the minimum degree of a vertex in $ H $, define $ g(H)=\frac{2k-l-3}{k(l+1)(m+2)} $. Since $ l \leq k-1 $, $ g(H) $ is always positive.

\begin{thm}
\label{thm:containment}
If H is a graph with $ k \ge 3$ vertices, then $ H $ containment has quantum query complexity $ O\left(n^{2-2/k-g(H)}\right) $.
\end{thm}

This result improves the quantum query complexity of $ H $ containment from  $ O\s{n^{2-2/k}} $ in Theorem \ref{thm:containmentold} to $ O\left(n^{2-2/k-g(H)}\right) $. Table \ref{tbl:subgraphcomplexity} summarises the query complexities for some typical subgraph containment using this general learning graph. If $ H $ is a bipartite graph, $ g(H) $ is slightly complicated and our result is usually worse than the result for bipartite graph containment in \cite{Childs11}.

\begin{table}[htb]
\centering
\begin{tabular}{|c|c|c|p{5cm}|}
\hline
Subgraph $ H $ & $ g(H) $ & complexity & remark\\
\hline
$ k $-clique & $ \frac{2(k-2)}{k^2(k^2-3k+6)} $ & $ O\s{n^{2-\frac{2}{k}-\frac{2(k-2)}{k^2(k^2-3k+6)}}} $ & best so far\\
\hline
path $ P_k $ & $ \frac{k-2}{k^2} $ & $ O\s{n^{2-\frac{3k-2}{k^2}}} $ & worse than \cite{Childs11}\\
\hline
cycle $ C_k $ & $ \frac{2k-5}{3k^2} $ & $ O\s{n^{2-\frac{8k-5}{3k^2}}} $ & best so far for $ k $ odd, worse than \cite{Childs11} for $ k $ even\\
\hline
star $ S_k $ & $ \frac{k-1}{(k+1)^2} $ & $ O\s{n^{2-\frac{3k+1}{(k+1)^2}}} $ & worse than \cite{Childs11}\\
\hline
\end{tabular}\caption{A summary of the query complexity for various subgraph containment problem using the learning graph described by Table \ref{tbl:subgraph}}
\label{tbl:subgraphcomplexity}
\end{table}

\pfstart
Here, a random subgraph $ U $ on $ k $ vertices is a graph on $ k $ vertices such that each edge is present with probability $ s $, independently at random. The set of $ k $ vertices is also taken uniformly at random. The random graph contains both the edges and the $ k $ vertices, though some of the vertices in $ U $ may have degree 0. Randomly querying a set of edges means querying each edge in the set with probability $ s $.
 
Denote the vertices of $ H $ by $ a_1, \cdots, a_k $, with $ a_k $ having degree $ l $. By deleting vertex $ a_k $ and all the edges connected to it, we get a subgraph $ G $ of $ H $. $ G $ has $ m $ edges. Denote  the set of edges $ e_i $ in $ G $ by $ M $. $ M=\{e_1, \cdots, e_m\} $. In Table \ref{tbl:subgraph} we give a learning graph for this problem.
\begin{table}[htb]
\centering
\begin{tabular}{rp{13cm}}
\hline
1. & Query a random subgraph on $ r-k+1 $ vertices that does not contain vertices $ a_1, \cdots, a_k $.\\
2. & Randomly query edges connecting $ a_1 $ to the vertices of the subgraph.\\
3. & Randomly query edges connecting $ a_2 $ to the vertices of the subgraph (including $ a_1 $).\\
$ \vdots $\\
$ k $. & Randomly query edges connecting $ a_{k-1} $ to the vertices of the subgraph, (including $ a_1, \cdots a_{k-2}) $.\\
$ k+1 $. & Select those L-vertices that do not contain any edge in $ M $ and contain at least $ sr^2/4 $ edges.\\
$ k+2 $. & Query $ M $.\\
$ k+3 $. & Use a subroutine from Table \ref{tbl:kdistinctness} to query the $ l $ edges out of all edges connecting $ a_k $ to the vertices of the subgraph.\\
\hline
\end{tabular}
\caption{Stages for the $ H $-finding algorithm.}
\label{tbl:subgraph}
\end{table}
\begin{table}[htb]
\centering
\begin{tabular}{|r|ccccccc|}
\hline
Stage & 1 & 2 & 3 & $ \cdots $ & $ k $ & $ k+2 $ & $ k+3 $\\
\hline
Speciality & 1 & $n$ & $n^2/r$ & $ \cdots $ &$ n^{k-1}/r^{k-2} $ & $ n^{k-1}/r^{k-3}s^{m-1} $ & $ n^k/r^{k-1}s^m $\\
Length/Complexity & $sr^2$ & $ sr $ & $ sr $ & $ \cdots $ & $ sr $ & 1 & $ r^{l/(l+1)} $\\
\hline
\end{tabular}\caption{Parameters (up to a constant factor) of the stages in the learning graph described by Table \ref{tbl:subgraph}}
\label{tbl:subgraphParam}
\end{table}

Stage $ k+1 $ is not a stage of transitions between layers. Rather, it modifies the flow in transitions in all of the previous stages. Denote the probability of the L-vertex of the constructed learning graph to satisfy the constraint of Stage $ k+1 $ by $ p $. Assuming $ s=o(1) $ and $ sr^2=\omega(1) $, the probability is $ 1-o(1) $. Assume the instance is large enough, so $ p\geq 1/2 $. Then we scale up the flow $ 1/p $ times and remove all flow going to the bad L-vertices. Since we are modifying the flows in all the previous stages, it is easy to see that the flow through each L-vertex is preserved. By Lemma \ref{thm:modifyflow}, the complexity of this stage remains the same, up to a constant factor. Therefore, it is sufficient to calculate the complexity of this stage before modification.

Before modification, we can think of the flow as the probability of getting a valid transition out of all valid transitions across that stage, since the invalid transitions have zero flow by definition. A transition is valid here if the subgraph being queried at stage 1 is correct and each transition at previous stages queries the correct edges. Ignoring the edges, the number of ways of getting the correct vertices at stage 1 is $ \binom{n-k}{r-k+1} $. For Stage j, the probability that the starting subgraph contains $ u $ edges, given that it's valid, is $ s^u(1-s)^{\binom{r-k+j-1}{2}-u} $. The probability that a transition in this stage queries $ v $ edges is $ s^v(1-s)^{r-k+j-1-v} $. So, if a valid transition in Stage $ j $ starts from a subgraph of $ u $ edges and queries $ v $ edges, the probability of getting it out of all valid transitions at Stage $ j $ is $ \binom{n-k}{r-k+1}^{-1}s^{u+v}(1-s)^{\binom{r-k+j}{2}-u-v} $. Another valid transition in this equivalent class would have the same flow, since $ u $ and $ v $ are the same, and this flow does not depend on the input. Using the same argument as those in section 4, it can be checked that the speciality of this transition is $ O\s{n^{j-1}/r^{j-2}} $. Unlike the previous cases, not all transitions belong to the same equivalence class. However, essentially the same argument shows the speciality of each equivalence class is the same. Hence, this stage satisfies the symmetry requirement. Similarly, it can be checked that all other stages satisfy the symmetry requirement.

\begin{table}[htb]
\centering
\begin{tabular}{|r|cccc|}
\hline
Substage & 1 & 2 & $ \cdots $ & $ m $ \\
\hline
Speciality & $ n^{k-1}/r^{k-3} $ & $n^{k-1}/r^{k-3}s^1$ & $ \cdots $ &$n^{k-1}/r^{k-3}s^{m-1} $ \\
Length/Complexity & 1 & 1 & $ \cdots $ & 1 \\
\hline
\end{tabular}\caption{Parameters (up to a constant factor) of Stage $ k+2 $ in the learning graph described by Table \ref{tbl:subgraph}}
\label{tbl:k+2Param}
\end{table}

Stage $ k+2 $ is a compilation of $ m $ substages, being \textit{query} $ e_1 $, \textit{query} $ e_2 , \cdots$ \textit{query} $ e_m $. For Substage $ i $, the probability a random permutation of vertices identifies the 2 vertices (e.g. $ a $ and $ b $) of edge being added correctly is exactly $ \frac{2}{n(n-1)} $. Provided that this happens, the probability that $ a_k $ is not used in the vertex set of the subgraph is $ (n-r)/(n-2) $. The probability that the vertex set includes all other vertices of $ a_1, \cdots a_{k-1}$ is $ O \left(r^{k-3}/n^{k-3}\right) $. But then $ ab $ is only connected with probability $ s $, and the same for the vertices identified in previous stages. So, provided the permutation identifies the vertices correctly, the probability that the edges added in all previous substages are among the edges of the subgraph for a fixed choice of the vertex set containing $ a_1, \cdots a_{k-1} $ is $ O\left(s^{i-1}\right) $. So the speciality of Substage $ i $ is $ O\left(n^{k-1}/r^{k-3}s^{i-1}\right) $. The speciality of Stage $ k+2 $ is $ O\left(n^{k-1}/r^{k-3}s^{m-1}\right) $.

By Theorem\ref{thm:summa}, the total complexity is \begin{center}
$ O\left(sr^2+sr\sum\limits_{i=1}^{k-1}\sqrt{\frac{n^i}{r^{i-1}}}+\sqrt{\frac{n^{k-1}}{r^{k-3}s^{m-1}}}+r^{l/(l+1)}\sqrt{\frac{n^k}{r^{k-1}s^m}}\right) $.
\end{center}

Since Stage $ k $ dominates all stages from 2 to $ k-1 $, the total complexity is simplified to $ O\left(sr^2+sr\sqrt{\frac{n^{k-1}}{r^{k-2}}}+\sqrt{\frac{n^{k-1}}{r^{k-3}s^{m-1}}}+r^{l/(l+1)}\sqrt{\frac{n^k}{r^{k-1}s^m}}\right) $. This is optimised to \begin{center}
$ O\left(n^{2-\frac{2}{k}-\frac{2k-l-3}{k(l+1)(m+2)}}\right) $
\end{center} when $ r=n^{1-1/k} $ and $ s=n^{-\frac{2k-l-3}{k(l+1)(m+2)}} $. Since $ \frac{2k-l-3}{k(l+1)(m+2)} $ increases as $ l $ decreases (while keeping $ m+l $ fixed), choosing $ a_k $ as the vertex of the smallest degree in Stage $ k+3 $ minimises the complexity in this model. Also, the assumptions $ s=o(1) $ and $ sr^2=\omega(1) $ are valid.
\pfend

Theorem \ref{thm:containment} can be naturally extended to monotone graph properties with small 1-certificates. A graph property is monotone (or monotone increasing) if it is preserved under the addition of edges and vertices (sometimes edges only, depending on the literature). If $ \phi $ is a monotone graph property whose 1-certificates have at most $ K>3 $ vertices, let $ \Phi $ be the set of 1-certificates $ H $. $ \left|{\Phi}\right|<G $ for some contant $ G $, which depends on $ K $ only. Let $ \tilde{g}\s{\phi}=\min_{H\in \Phi}\s{2/k\s{H}+g\s{H}} $, where $ k\s{H} $ denotes the number of vertices in $ H $. To check if a graph has monotone graph property $ \phi $, we can just apply Theorem \ref{thm:containment} to all its 1-certificates $ H $.
\begin{coly}
Let $ \phi $ be a monotone graph property whose 1-certificate have bounded size. Then checking $ \phi $ and producing a 1-certificate $ H $ when $ \phi $ is satisfied, can be done with quantum query complexity $ O\s{n^{2-\tilde{g}\s{\phi}}} $.
\label{thm:monotone}
\end{coly}

\section{Conclusion and open problems}
We give an improved quantum query complexity for subgraph containment problem. This shows that the learning graph model is indeed very powerful. However, our results do not imply an improved time complexity, which is also important in quantum algorithms.

We observed that the learning graph model could recover some of the results previously obtained using quantum walks on the Johnson graph. However, it is not yet known if it can be used to recover recent results obtained using quantum walks on the Hamming graph \cite{Childs11}, especially on $ C_4 $ containment and path finding. Also, it might be interesting to get a more intuitive understanding about the learning graph model and construct explicit quantum walk algorithms from the learning graph with the same quantum query complexity.

Recently, a modified learning graph model was proposed by Belovs and Lee \cite{Belovs11a}. This model seems to have more flexibility. Naturally, one would wonder if this model could further improve the results, and possibly include problems with non-constant 1-certificate complexity.

{\bf Note added.} Following the completion of this work, I became aware of recent independent work by Lee, Magniez and Santha \cite{Lee11}. They obtained the same quantum query complexity for constant-sized subgraph containment problem in the learning graph model, using a different technique.

\subsection*{Acknowledgement}
I would like to thank Richard Jozsa and Ashley Montanaro for many helpful conversations and for feedback on early versions of this work. I would like to thank Aleksandrs Belovs and Andrew Childs for correcting some mistakes in my paper. I would also like to thank Miklos Santha for bringing the work \cite{Lee11} to my attention.

This work was supported in part by Queens' College and DAMTP.

\end{document}